# Hybrid integration of silicon photonics circuits and InP lasers by photonic wire bonding

Muhammad Rodlin Billah,[1,2] Matthias Blaicher,[1,2] Tobias Hoose,[1,2] Philipp-Immanuel Dietrich,[1,2,3] Pablo Marin-Palomo,[2] Nicole Lindenmann,[1,2] Aleksandar Nesic,[2] Andreas Hofmann,[4] Ute Troppenz,[5] Martin Moehrle,[5] Sebastian Randel,[2] Wolfgang Freude,[2] and Christian Koos[1,2,3,*]

[1]Institute of Microstructure Technology (IMT), Karlsruhe Institute of Technology (KIT), Hermann-von-Helmholtz-Platz 1, 76344 Eggenstein-Leopoldshafen, Germany
[2]Institute of Photonics and Quantum Electronics (IPQ), Karlsruhe Institute of Technology (KIT) Engesserstrasse 5, 76131 Karlsruhe, Germany
[3]Vanguard Photonics GmbH, Karlsruhe, Germany
[4]Institute of Applied Computer Science (IAI), Karlsruhe Institute of Technology (KIT), Hermann-von-Helmholtz-Platz 1, 76344 Eggenstein-Leopoldshafen, Germany
[5]Fraunhofer Institute for Telecommunications, Heinrich Hertz Institute (HHI), Einsteinufer 37, 10587 Berlin, Germany
*Corresponding author: christian.koos@kit.edu

**Abstract:** Efficient coupling of III-V light sources to silicon photonic circuits is one of the key challenges of integrated optics. Important requirements are low coupling losses, as well as small footprint and high yield of the overall assembly, along with the ability to use automated processes for large-scale production. In this paper, we demonstrate that photonic wire bonding addresses these challenges by exploiting direct-write two-photon lithography for in-situ fabrication of three-dimensional freeform waveguides between optical chips. In a series proof-of-concept experiments, we connect InP-based horizontal-cavity surface emitting lasers (HCSEL) to passive silicon photonic circuits with insertion losses down to 0.4 dB. To the best of our knowledge, this is the most efficient interface between an InP light source and a silicon photonic chip that has so far been demonstrated. Our experiments represent a key step in advancing photonic wire bonding to a universal integration platform for hybrid photonic multi-chip assemblies that combine known-good dies of different materials to high-performance hybrid multi-chip modules.

## 1. INTRODUCTION

The silicon photonic platform has evolved into a mainstay of high-density photonic integration [1], exploiting advanced CMOS processes for high-yield mass fabrication [2,3] of a wide variety of passive photonic devices [4,5], electro-optic modulators [6,7], or photodetectors [8,9]. Cost-efficient and technically viable integration of high-performance light sources into silicon photonic circuits, however, still remains a challenge [10]. Such light sources should combine low power consumption and high efficiency with small footprint and good thermal coupling to package-level heat-sinks while still maintaining good manufacturability and the potential for automated large-scale production. Moreover, as system complexity increases, testing of active components prior to integration into the final system is becoming more and more important to achieve high fabrication yield. While substantial progress has been made towards realizing light sources by direct epitaxial [11,12] or heteroepitaxial [13–15] growth of direct-bandgap compound semiconductors on silicon substrates, wafer-level [16–18] or die-level [19–21] transfer of III-V layer stacks or even complete devices [22–26] to silicon substrates is still considered the most practical path towards an efficient light source integration. In this context, two main approaches have been pursued, which are often referred to as heterogeneous integration and hybrid integration [27–30].

Heterogenous integration is based on bonding of unpatterned III–V dies onto pre-processed silicon photonic (SiP) wafers such that light generated in the epitaxial layers of the III–V die is evanescently coupled to the SiP waveguides [16–21]. III-V devices are then fabricated by wafer-scale processing of the dies, where all structures are lithographically aligned with highest precision. High-precision positioning of dies is hence not necessary – a key advantage compared to hybrid integration. However, while heterogeneous integration is particularly well suited for large-scale mass fabrication of III-V light sources on SiP circuits, the associated technical complexity is still considerable, in particular with respect to the stringent requirements of ultraclean and extremely smooth surfaces. Moreover, heterogeneous integration does not permit testing of light sources prior to integration into more complex systems and hence requires tight process control to maintain high yield. As a consequence, heterogeneous integration is mainly suited for high-volume applications that justify associated technological overhead, e.g., in the context of optical on-chip networks that require integration of tens of light sources onto a single chip [31]. Moreover, heterogeneously integrated light sources may consume considerable real estate on the SiP chip, and heat-sinking is challenging due to the high thermal resistance introduced by the III-V-to-silicon bonding layer and by the buried oxide [32].

Hybrid integration [22–26], in contrast, relies on optically connecting readily processed III-V lasers [22–24], gain chips [25,26], or even photodiodes [27] to silicon photonic (SiP) circuits, where the III-V device may either be mounted on top of the silicon substrate [22–25,27] or next to it [26]. Hybrid integration maintains the superior performance characteristics of native III-V light sources and allows for testing of devices prior to system assembly, but comes with fabrication challenges. In particular, efficient optical coupling of the III-V to the SiP waveguides usually relies on alignment with precisions in the lower micrometer or even sub-µm range. This often requires slow and expensive [33] active alignment techniques, where the coupling efficiency is continuously monitored while optimizing the position of the devices [10,34]. Moreover, additional devices such as micro-lenses, prisms, or micro-mechanical carriers are needed to adapt the mode field size and the emission direction of the III-V light source to that of the SiP circuit [22], leading to comparatively big assemblies. In many cases, the III-V devices are mounted on top of the SiP die. This approach does not only consume substantial on-chip real estate, but also poses challenges with respect to heat-sinking of the III-V devices through the underlying silicon-on-insulator (SOI) substrate due to the relatively poor heat conductivity of the buried oxide [23,28,35]. In hybrid integration, optical coupling losses of 2.3 dB have been previously demonstrated for butt coupling of a III-V laser diode array to an array of SiP waveguides equipped with trident spot-size converters [36] that relax alignment tolerances to ± 0.7 µm.

In this paper, we demonstrate that the technique of photonic wire bonding [37–41] enables highly flexible low-loss coupling of InP light sources to SiP chips, maintaining all performance and flexibility advantages of hybrid integration approaches while offering a path towards highly scalable automated mass production. Photonic wire bonding exploits in-situ additive nanofabrication of freeform polymer waveguides between pre-positioned photonic chips. The three-dimensional shape of the photonic wire bonds can be adapted to the exact positions of the chips such that high-precision alignment of chips becomes obsolete, rendering the technique amenable to automated mass production. Building upon our previous results [37–41], we demonstrate highly efficient coupling between InP-based horizontal-cavity surface-emitting lasers (HCSEL) and silicon photonic chips with coupling losses down to 0.4 dB. In these assemblies, the laser source is placed side-by-side to the SiP chip, thus allowing for good thermal coupling to an underlying chip-level heat sink without consuming any on-chip real estate. We experimentally confirm that the emission performance of the lasers is not impaired by the photonic wire bonding process. Combined with previously demonstrated chip-to-chip [39] and fiber-to-chip [40] connections, our experiments represent a key step in advancing photonic wire bonding to a universal integration platform for hybrid photonic multi-chip assemblies that combine known-good dies of different materials while maintaining their individual high-performance characteristics.

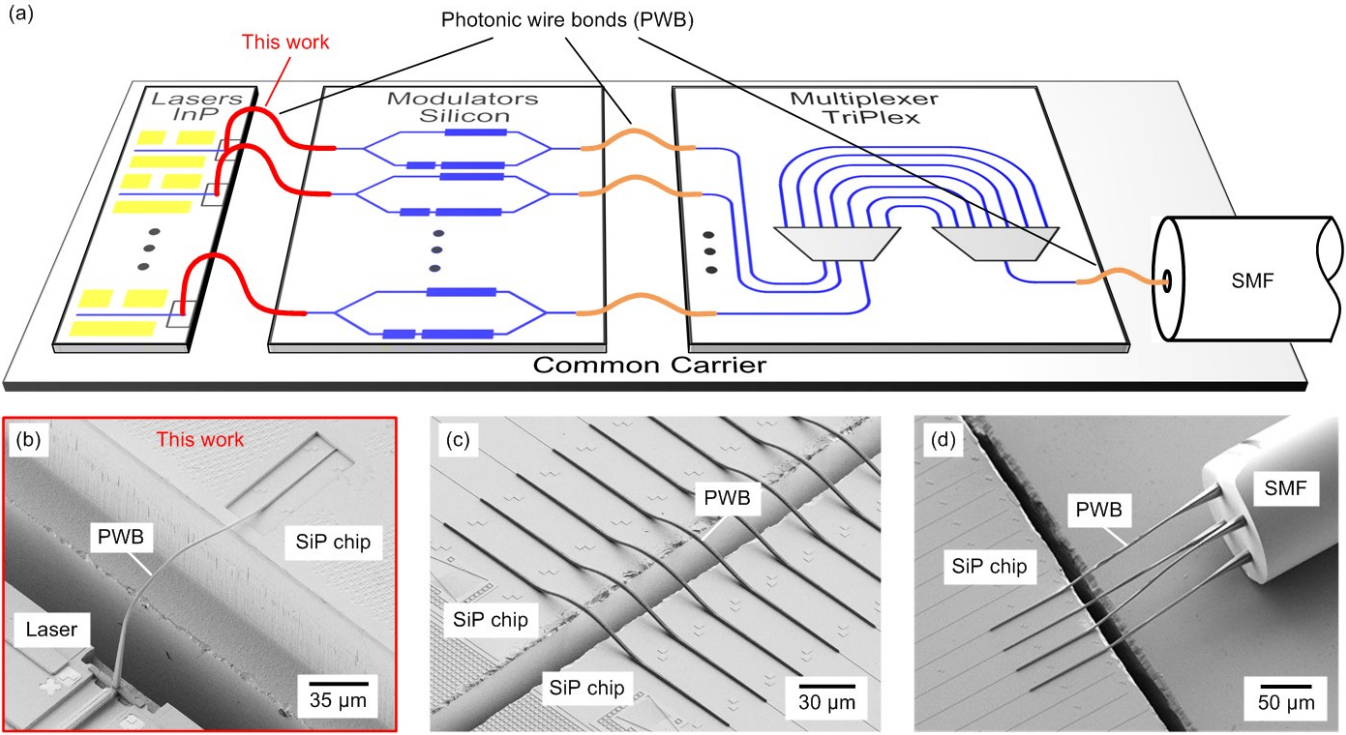

**Fig. 1.** Vision of a photonic multi-chip transmitter for wavelength-division multiplexing (WDM) communications. **(a)** The system exploits photonic wire bonds to combine the distinct advantages of different photonic integration platforms: Distributed-feedback (DFB) lasers based on direct-bandgap InP-substrates, whereas silicon photonic (SiP) chips lend themselves for implementing electro-optic modulators. For dense packing of optical channels, high-quality arrayed waveguide gratings (AWG) are needed, which are best realized on medium index-contrast material systems such as TriPleX [42]. The functionality of the multi-chip system depends vitally on efficient chip-to-chip and chip-to-fiber interconnects, which are realized by photonic wire bonds (PWB). The focus of this work is on low-loss PWB links between InP lasers and SiP chips (red colored). **(b)** SEM image of a laser-chip interface. **(c)** Chip-to-chip-connection between two SiP dies [39]. **(d)** Fiber-to-chip interface using PWB to connect the individual cores of a multi-core fiber (MCF) to an array of planar SOI waveguides [40].

## 2. CONCEPT AND BACKGROUND

A vision of a hybrid photonic multi-chip module enabled by photonic wire bonds is illustrated in Fig. 1(a) using a wavelength-division multiplexing (WDM) transmitter as an example. The system combines the distinct advantages of different photonic integration platforms: Distributed-feedback (DFB) lasers are used as optical sources for the various wavelength channels and implemented on indium phosphide (InP) substrates, whereas silicon photonics (SiP) chips are used to realize densely integrated electro-optic IQ modulators which encode information on the various optical carriers. For dense packing of WDM channels, high-quality optical filters, e. g., arrayed waveguide gratings (AWG), are needed. These devices are best realized on the basis of a medium index-contrast material system such as the commercial integration platform TriPleX [42]. Hybrid multi-chip integration allows on-chip subsystems to be tested prior to integration which significantly increases yield as compared to monolithic or heterogeneous integration approaches, where failure of a single component impairs the functionality of the entire system. In previous experiments [39,40], we have shown photonic wire bonds to enable particularly efficient single-mode chip-to-chip and fiber-to-chip connections, Fig. 1(c) and (d), and the viability of the concept has been recently demonstrated by realizing multi-chip transmitter modules for high-speed communications [43,44]. The focus of this paper is on low-loss coupling of InP-based light sources to SiP chips as one of the most important interfaces for practical applications, Fig. 1(b).

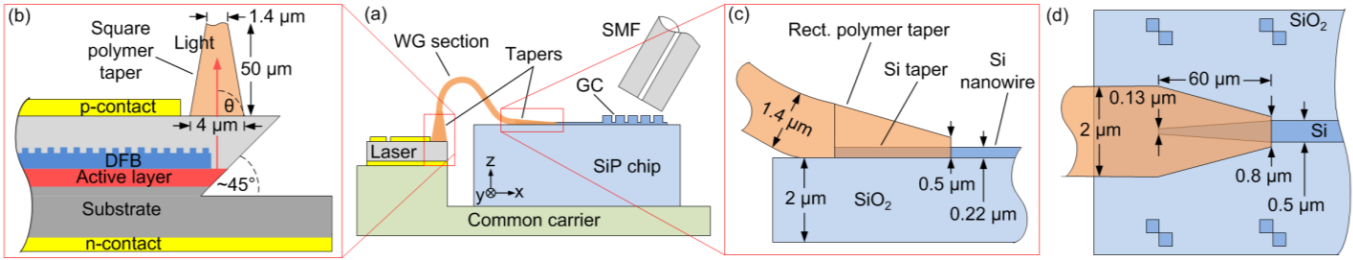

**Fig. 2.** Multi-chip assembly demonstrating a PWB between an InP laser and a SiP chip. **(a)** Assembly concept: The laser and the SiP die are mounted on a common carrier by adhesive bonding. The carrier compensates differences of die thickness. Precise alignment between the laser emission window and the Si nanowire is not required. Tapers are utilized at the interfaces at both ends of the PWB for a low-loss transition between the PWB waveguide (WG) section and the attached components. A grating coupler (GC) is used to interface the SiP waveguide to a standard single-mode fiber (SMF). The WG section is designed to have a rectangular cross-section of 2.0 µm × 1.4 µm. **(b)** HCSEL interface: The HCSEL consists of an in-plane InGasAsP DFB laser cavity and an etched 45° mirror to deflect the light to the surface-normal direction. The light is then captured by a (rectangular) polymer PWB taper. At the HCSEL emission window, the taper cross section corresponds to a square with a side length of 4 µm, which is linearly converted to a rectangular cross section with size of 2.0 µm × 1.4 µm at the transition to the PWB waveguide section. **(c, d)** Side and top view of the transition to the SiP nanowire waveguide. The polymer PWB taper starts with a rectangular 2.0 µm × 1.4 µm cross section of the PWB waveguide section, which is linearly converted to a final width of 0.8 µm and a height of 0.5 µm along a length of 60 µm. Alignment markers are used to exactly locate the coupling interface to the position of the targeted nanowire.

## 3. REALIZATION OF MULTI-CHIP MODULES

### A. Assembly

In our experiment, we realize simple multi-chip modules (MCM) that use photonic wire bonds (PWB) to connect passive SiP circuits to InP light sources, see Fig. 2(a). The assembly of the MCM requires a common carrier where laser and SiP chips are mounted first with medium precision by adhesive bonding. Height differences between the chips are compensated by the carrier to roughly align the top surfaces. The output port of the InP chip is then interconnected to a silicon (Si) nanowire on the SiP chip by a 3D freeform PWB. At the end of the Si nanowire, a grating coupler (GC) is used to direct the light out of plane into a single-mode fiber (SMF). Reference nanowires with GC at both ends are also located on the same SiP chip for calibration. The PWB is fabricated in-situ by two-photon lithography [39] and comprises two tapered sections to minimize mode-field mismatch at both interfaces. The shape of the PWB is adapted to the position of these interfaces, thereby compensating tolerances in chip placement.

### B. Photoresist materials and devices

Photonic wire bonds are fabricated from a commercially available negative-tone photoresist (IP-Dip, Nanoscribe GmbH, unexposed refractive index $n_{PWB}$ = 1.52 at 780 nm [45]). The MCM are built with so-called horizontal-cavity surface-emitting lasers (HCSEL) [46] which combine an in-plane InGaAsP-based buried heterostructure distributed-feedback (DFB) laser with an etched 45° mirror to deflect the emitted light to surface-normal direction, Fig. 2(b). At the HCSEL facet, the $1/e^2$-diameter of the modal intensity profile amounts to 3 µm, extracted from a measurement of the far-field intensity distribution by using a scanning-slit optical beam profiler (BP209-IR/M, Thorlabs). The HCSEL deflection mirror is fabricated by chemically assisted ion beam etching [46]. Note that the mirror inclination is subject to fabrication tolerances such that the light emission direction might not be perpendicular to the chip surface, i.e., θ ≠ 90° in Fig. 2(b).

The PWB consists of a waveguide (WG) section, which is connected to the HCSEL and the SiP chip by dedicated tapers, Fig. 2(a). On the HCSEL side, a taper with a rectangular cross section is used to convert the mode of the laser to the fundamental mode of the PWB waveguide section. At the HCSEL emission window, Fig. 2(b), the taper cross section corresponds to a square with a side length of 4 µm, which is then linearly converted into a rectangular cross section of 2.0 µm × 1.4 µm at the transition to the PWB waveguide section. At the interfaces to the SiP circuit, low-loss coupling is achieved by a tapered Si nanowire, which is embedded into the polymer PWB taper [39], see Fig. 2(c, d). The Si nanowires ($n_{Si}$ = 3.48) were produced on a CMOS pilot line using 193 nm deep-ultraviolet (DUV) lithography with standard height of 220 nm. The thickness of buried oxide (SiO$_2$, $n_{SiO2}$ = 1.44) amounts to 2 µm. The Si nanowire starts with a tip width of 0.13 µm, which then is tapered up to the final SiP waveguide width of 0.5 µm along a length of 60 µm. The polymer PWB taper starts with the rectangular 2.0 µm × 1.4 µm cross section of the PWB waveguide section, which is linearly converted to a final width of 0.8 µm and a height of 0.5 µm along the same length of 60 µm. The smallest radius of curvature typically used in the WG section of the PWB is 40 µm.

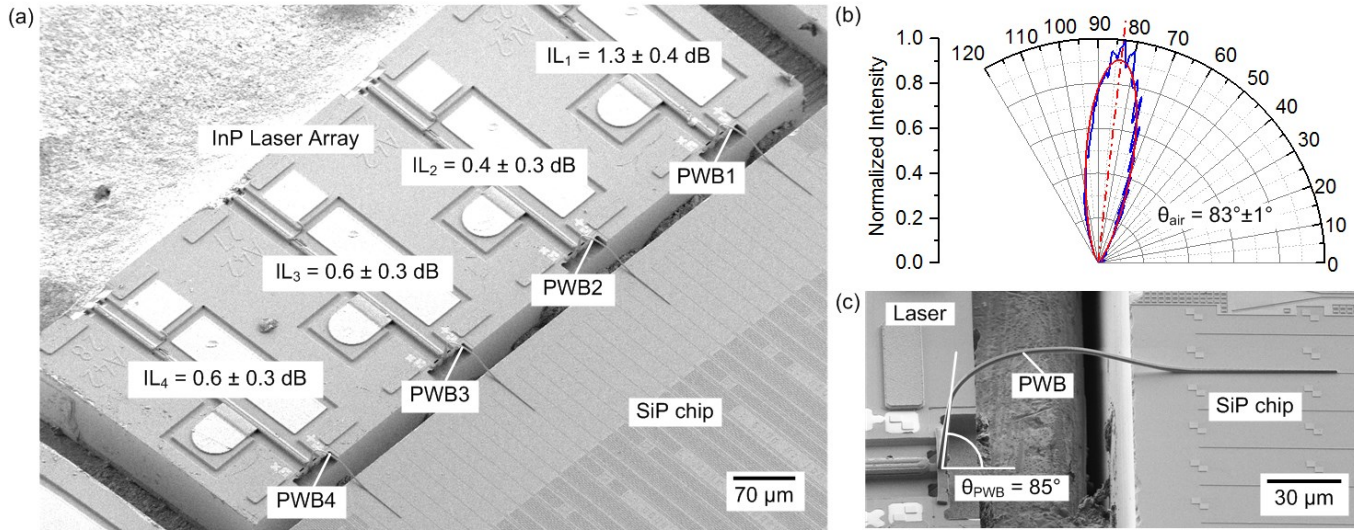

**Fig. 3.** Hybrid MCM combining passive silicon photonic (SiP) waveguides with an InP DFB laser array. **(a)** Scanning electron microscope (SEM) micrograph of the MCM, comprising four laser diodes (LD1 … LD4), each connected to a SiP on-chip waveguide via a photonic wire bond (PWB1 … PWB4). The insertion losses of the photonic wire bonds are denoted in the figure. PWB2 shows an insertion loss of $(0.4 \pm 0.3)$ dB, including the coupling losses of both the HCSEL-PWB interface and the transition to the SiP nanowire as well as the propagation loss in the freeform waveguide. Slightly higher losses of $(1.3 \pm 0.4)$ dB, $(0.6 \pm 0.3)$ dB, and $(0.6 \pm 0.3)$ dB are observed for PWB 1, 3, and 4, respectively. The variations are attributed to uncertainties of the HCSEL emission spot size, of the emission direction, and of the PWB shape. Still, these losses are well below the 2.3 dB obtained for other concepts that rely on active alignment [36]. **(b)** Measurement of emission direction of a HCSEL before photonic wire bonding. The measurement was taken by using a goniometric radiometer for three HCSEL chips. The Gaussian fit (red) reveals an emission angle $\theta_{air} = (83 \pm 1)°$, which deviates from the ideal 90° due to fabrication tolerances of the HCSEL deflection. **(c)** Side view of PWB1. For efficient coupling into the PWB taper ($n_{PWB} = 1.52$), the taper axis has to be inclined by approximately $\theta_{PWB} = 85°$ according to Snell's law.

### C. Photonic wire bond fabrication

Several steps are required for fabricating the PWB [37–40]. After mounting the components on a common carrier with typical accuracies of 10 µm, the negative-tone photoresist is drop-cast onto the MCM, covering both the HCSEL emission window and the coupling region of the targeted Si nanowire. The position of the coupling interfaces is then detected in the volume of the resist through the observation camera of the 3D lithography system (Photonic Professional GT, Nanoscribe GmbH). These positions may be derived from dedicated alignment markers, see Fig. 2(d). The start and end points of the PWB allow to define its trajectory along with its 3D shape, and the structure is then fabricated through two-photon polymerization by exposing the photoresist. The lithography laser has an emission wavelength of $\lambda = 780$ nm and emits pulses with a full-width at half-maximum (FWHM) below 100 fs and with a repetition frequency of 80 MHz. In our current experiments, we have not yet optimized the writing speed, leading to an exposure time of approximately three minutes for a single PWB with vast potential for further acceleration. The unexposed photoresist is finally removed in a two-step development process using propylene-glycol-methyl-ether-acetate (PGMEA) for the first 15 minutes followed by isopropyl-alcohol (2-propanol) inside a critical point dryer (CPD 300, Leica Microsystems GmbH) for 50 minutes. Low-refractive index matching liquid (Cargille Laser Liquid 3421, $n_{oil} = 1.3$ at 1550 nm) is drop-cast onto the assembly to emulate the low-refractive-index cladding of the PWB. The liquid can easily be replaced with a long-term protective cladding material.

## 4. RESULTS AND DISCUSSION

### A. Photonic wire bonding of laser array

To demonstrate the viability of the photonic wire bonding approach, we fabricate a hybrid MCM that combines passive silicon photonic waveguides with an array of DFB lasers that feature different wavelengths. An SEM micrograph of the assembly is shown in Fig. 3(a). Note that due to fabrication tolerances of the HCSEL deflection mirror inclination, the angle θ between the light emission direction and the chip surface is not exactly 90°, and hence the direction of the PWB trajectory has to be adapted accordingly, see Fig. 3(b, c). Using a goniometric radiometer (LD 8900, Ophir Spiricon Europe GmbH), we find that actual emission angle amounts to $\theta_{air} = (83 \pm 1)°$ when measured in air ($n_{air} = 1$), Fig. 3(b). For coupling into the polymer taper of the PWB ($n_{PWB} = 1.52$), this changes to approximately $\theta_{PWB} = 85°$ according to Snell's law, and the axes of the polymer tapers have to be adapted accordingly, Fig. 3(c).

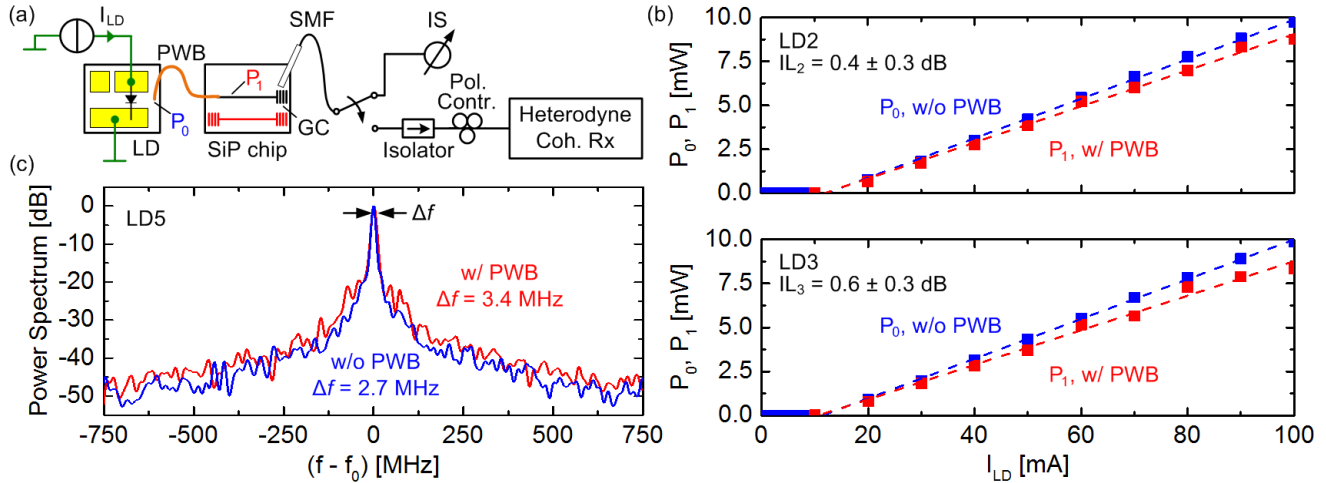

**Fig. 4.** PWB performance characterizations. **(a)** Experimental setup for measuring insertion losses of the PWB as well as laser linewidth. The electrical path (green) consists of a current source to drive the laser diode (LD). The optical path comprises the PWB (orange) as well as a SiP nanowires grating coupler (GC) and an optical single-mode fiber (all black). At the end of the single-mode fiber (SMF), the light is coupled into a calibrated integrating sphere (IS) to precisely measure the optical power. Alternatively, the emitted light can be coupled to a heterodyne coherent receiver, which contains a highly stable external-cavity laser (ECL, not depicted) as a local oscillator for linewidth measurements. A polarization controller (Pol. Contr.) is used for optimizing the beat signal between the recorded light and the LO reference. **(b)** Power-current characteristic of LD2 and LD3 measured at the laser facet before photonic wire bonding ($P_0$, blue squares), and inside the respective SiP nanowire after photonic wire bonding ($P_1$, red squares). The power level $P_1$ was corrected by taking into account the losses caused by propagation through the SiP chip, the corresponding GC, and the SMF patch cord. The dashed lines represent linear fits. The PWB insertion losses amount to 0.4 dB and to 0.6 dB respectively. **(c)** Optical linewidth of LD5 measured before and after photonic wire bonding. The full-width at half-maximum of the power spectrum amounts to $\Delta f = 2.7$ MHz without and $\Delta f = 3.4$ MHz with a PWB. Within the measurement accuracy, we do not see a significant increase of the laser linewidth. These findings were confirmed using a variety of other devices.

### B. Insertion loss of PWB

In order to precisely measure the insertion loss (IL) of each PWB, we first determine the current-dependent output power at each HCSEL facet before and after photonic wire bonding, see Fig. 4(a). This is done by electrically pumping each laser diode (LD) with an adjustable current $I_{LD}$, derived from a laser diode driver (LDX 3620, ILX Lightwave). In a first measurement, we record the total output power of the bare HCSEL before photonic wire bonding. To this end, we bring an integrating sphere (IS, S145C, Thorlabs) in close proximity to the HCSEL facet such that the entire emitted power $P_0$ is captured. In a second measurement, which we perform after photonic wire bonding, we use the same integrating sphere to determine the optical power after propagation through the SiP chip, the corresponding grating coupler (GC) and a single-mode fiber (SMF) patch cord. In this measurement, we simply insert the SMF end facet into the IS to measure the power. For this measurement, we also consider the 4 % back reflection at the SMF end facet. Taking into account the on-chip propagation losses of the 472 µm-long SiP nanowire, the GC losses, and the connector losses of the SMF, we can then estimate the power $P_1$ delivered to the SiP nanowire. The wavelength-dependent GC losses and the on-chip propagation losses are obtained from reference SiP waveguides on the same chip. The transmission spectra of these waveguides are measured using a tunable laser and a synchronously swept optical spectrum analyzer (OSA). We investigate waveguides of different lengths, leading to on-chip propagation losses of approximately 3 dB/cm and grating coupler losses of 4.1 dB for the optimum wavelength of 1550 nm. For calculating the PWB insertion losses, we use the GC insertion loss at the emission wavelength of the respective HCSEL.

Figure 4(b) shows the measurement results for the emitted HCSEL power $P_0$ (blue, w/o PWB) and $P_1$ (red, w/ PWB) for LD2 and LD3. From the ratio of $P_0$ and $P_1$, we estimate the PWB IL to be (0.4 ± 0.3) dB and (0.6 ± 0.3) dB, respectively. Comparing the thresholds from both curves with and without PWB in Fig. 4(b), we confirm that the HCSEL thresholds stay at 10 mA before and after photonic wire bonding. Using the same method, we estimate the remaining PWB IL of LD1 and LD4 to be (1.3 ± 0.3) dB and (0.6 ± 0.3) dB, respectively, see Fig. 3(a). The IL differences are attributed to uncertainties of the HCSEL emission spot size and the emission direction as well as to slight variations of the PWB shape. Note that these losses are well below the 2.3 dB, which were obtained for other concepts that rely on active alignment [36].

### C. Laser performance without and with PWB

When connecting InP lasers to SiP chips, the use of intermediate optical isolators is not practical. An important question is then whether back-reflection of optical power into the laser cavity will degrade the emission performance, in particular with respect to the optical linewidth. This is investigated by first measuring the emission spectra to confirm that the lasers continue to emit in a single longitudinal mode after wire bonding. In a second step, we measure and compare the linewidth of the HCSEL without and with PWB. The laser linewidth measurement relies on superimposing the emitted LD light with light from a highly stable external-cavity reference laser acting as a local oscillator (LO). The beat signal is then detected using a heterodyne coherent receiver (N4391A, Keysight Technologies GmbH), see Fig. 4(a) and recorded by an oscilloscope (not shown). We record the intermediate-frequency signal, evaluate the corresponding variance of the phase increments and retrieve the Lorentzian linewidth [47]. The LO linewidth amounts to 14 kHz and can safely be neglected compared to the linewidth of the HCSEL, which is specified between 3 MHz and 5 MHz.

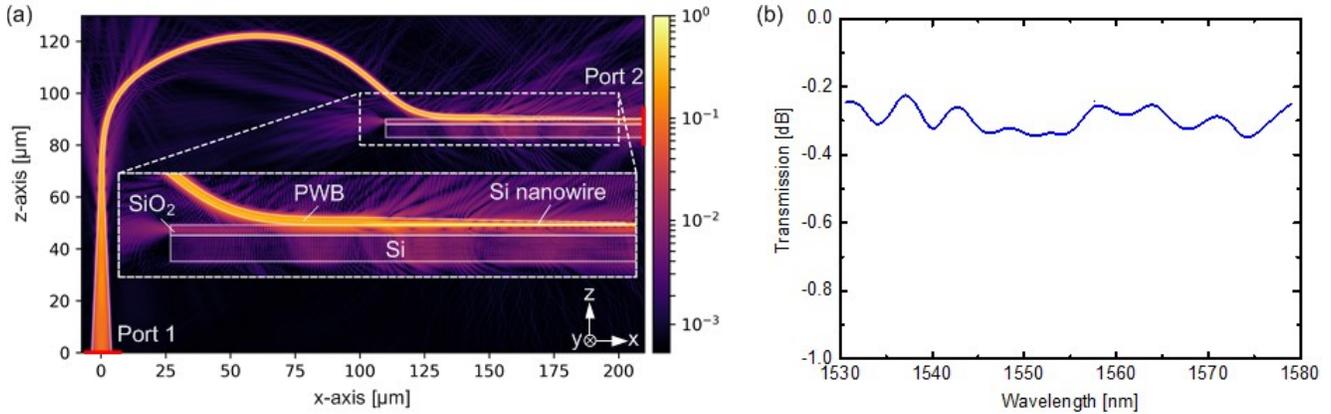

**Fig. 5.** Numerical verification and benchmarking of the PWB. **(a)** Calculated normalized intensity of the PWB. Simulations were performed using a vectorial finite-integration technique (FIT, CST Microwave Studio [48]). Light is launched into Port 1, which is located at the bottom of the PWB input taper connected to the HCSEL, and Port 2 extracts the power guided by the fundamental mode of the SiP waveguide. Losses are mainly caused by the transition between the PWB and the Si nanowire, indicated by field portions that are radiated away from the waveguide structure, see inset. **(b)** Transmission spectrum obtained for the propagation from Port 1 to Port 2. The insertion loss amounts to approximately 0.3 dB and is flat over the entire frequency range. This is in good agreement with the measured insertion losses that range from (0.4 ± 0.3) dB to (1.3 ± 0.4) dB.

Figure 4(c) shows the power spectrum of the beat signal obtained from another LD (LD5) without (blue) and with PWB (red). The optical linewidth $\Delta f$ of this laser can be inferred from the variance of the phase increments for small time delays and amounts to 2.7 MHz w/o and 3.4 MHz with PWB. This measurement was repeated for the other three devices (LD6 … LD8), leading to comparable results – measured linewidths changed from (2.9, 5.8, 4.2) MHz without PWB to (3.5, 3.1, 5.7) MHz with PWB. Note that the linewidth measurements of LD5 … LD8 were obtained from a previous MCM generation, in which the PWB IL amounted to 4 dB since the erroneous inclination of the HCSEL deflection mirror was yet not taken into account. This insertion loss is significantly higher than that of the low-loss MCM depicted in Fig. 3(a) for which we missed to measure the linewidth prior to photonic wire bonding. However, to confirm that the laser performance is also maintained for low-loss bonds, we also measure the linewidth of LD1 … LD4 after photonic wire bonding, leading to values of (1.9, 3.4, 2.8, 2.3) MHz. All these values are in the same range as the linewidths obtained for LD5 … LD8, which were fabricated on the same wafer. Based on these results, we conclude that the HCSEL performance is not altered by photonic wire bonding.

### D. Numerical verification of measurements

To benchmark and support our experimental results, we also performed a simulation of a complete PWB structure including the transition to the Si nanowire using a commercially available numerical solver (CST Microwave Studio [48]). We consider a simple case of a PWB having a fully planar trajectory, Fig. 5(a). Note that, in general, the positioning of the HCSEL chip with respect to the SiP chip is subject to mechanical tolerances that might, e. g., lead to a lateral offset. In this case, the axes of the HCSEL emission beam and of the Si nanowire waveguide do not lie in the same plane, and the PWB trajectory is non-planar.

In the simulation, we use the refractive indices and the device dimensions specified in the previous section along with a cladding of refractive index $n_{oil}$ = 1.3. We consider a frequency range between 190 THz and 196 THz. Waves are launched and detected at two simulation ports, which are located at the bottom of the PWB input taper connected to the HCSEL and at the end of Si nanowire. The ports are marked with red lines in Fig. 5(a). The field launched at Port 1 corresponds to the fundamental mode calculated for the starting cross section of the HCSEL taper. At Port 2, the power guided by the fundamental mode of the SiP waveguide is extracted to determine the coupling efficiency. The polarization of the launch field was chosen along the y-direction, leading to the excitation of a quasi-TE mode in the SiP nanowire waveguide. Note that this simulation does not take into account any scattering loss caused by surface roughness or any mode mismatch between the mode field at the HCSEL facet and the launched excitation field. Using the measured 1/e² diameter of 3 µm for the rotationally intensity profile of the HCSEL emission, we find a loss of only 0.03 dB for the transition to the square PWB input taper.

The plot in Fig. 5(a) shows the normalized intensity distribution obtained from the simulation at a frequency of 193 THz. The main losses originate from the transition of the PWB to the Si nanowire, indicated by radiated field patterns that propagate away from the Si nanowire, see inset. In contrast to that, the bends of the PWB WG section do not cause much loss. The simulated IL from Port 1 to Port 2 amounts to approximately 0.3 dB over the entire frequency range, see Fig. 5(b). The simulated losses are slightly smaller than their measured counterparts, which range from (0.4 ± 0.3) dB to (1.3 ± 0.4) dB.

## 5. SUMMARY

We have demonstrated that photonic wire bonding enables highly efficient coupling between InP-based horizontal-cavity surface-emitting lasers (HCSEL) and silicon photonic chips. We achieve very low coupling losses down to (0.4 ± 0.3) dB, which are in very good agreement with numerical simulations. To the best of our knowledge, this is the most efficient interface between an InP light source and a SiP chip that has so far been demonstrated. We further confirm experimentally that no detrimental effect on the DFB laser linewidth is introduced by the photonic wire bond. The photonic wire bonding approach enables flexible hybrid integration of best-in-class known-good devices with high density in a fully automated fabrication process. The technique lends itself to both rapid prototyping of small batch sizes and to fully automated large-scale production. Combined with previously demonstrated chip-to-chip [39] and fiber-to-chip [40] connections, we expect photonic wire bonding to evolve in a universal integration platform for hybrid photonic multi-chip assemblies.

**Funding**. The work was supported by BMBF joint projects PRIMA (grant 13N14630) and PHOIBOS (grant 13N12574), the H2020 Photonic Packaging Pilot Line PIXAPP (grant 731954), the European Research Council (ERC Starting Grant 'EnTeraPIC' 280145), the EU-FP7 project BigPIPES, the Alfried Krupp von Bohlen und Halbach Foundation, the Helmholtz International Research School for Teratronics (HIRST), the Karlsruhe School of Optics & Photonics (KSOP), and Karlsruhe Nano-Micro Facility (KNMF).

**Acknowledgment**. We acknowledge support by the Open Access Publishing Fund of Karlsruhe Institute of Technology (KIT), Florian Rupp for the SEM micrographs, Marco Hummel for the mechanical carrier, and Oswald Speck for the electric wire bonds.